\documentclass{aa}
\usepackage{graphicx}

\begin{document}

\newcommand{\fe}{\ion{Fe}{II}}
\newcommand{\h}{H$_2$}
\newcommand{\kms}{km\,s$^{-1}$ }
\newcommand{\um}{$\mu$m}
\newcommand{\brg}{Br$\gamma$}

\title{Observations of the Brackett decrement in the Class I source
\object{HH100 IR}
\thanks{Based on observations collected at the European Southern 
Observatory, Chile (ESO N.069.C-0269)}}
\author{B. Nisini\inst{1}, S. Antoniucci\inst{2,1}, T. Giannini\inst{1}}
\offprints{Brunella Nisini, nisini@mporzio.astro.it}

\institute{INAF-Osservatorio Astronomico di Roma, I-00040 Monteporzio Catone,
Italy \and Universit\`a degli Studi `Tor Vergata', via della Ricerca Scientifica
1, 00133 Roma, Italy
}
%
%
\date{Received 24 September 2003; Accepted 3 March 2004}
%
%
%
\titlerunning{HI lines in Class I sources}
\authorrunning{B. Nisini et al.}

\abstract{The Brackett decrement in the Class I source \object{HH100 IR}
has been observed and analyzed to set constraints on the 
origin of the IR HI emission in this young object. 
We have used both low resolution (R$\sim$800) observations of the
Brackett lines from \brg\, to Br24, and medium resolution (R$\sim$9000)
spectra of the \brg, Br12 and Br13 lines. The dereddened fluxes
indicates that the lines remain moderately thick up to 
high quantum numbers. Moreover, the profiles of the three
lines observed in medium resolution are all broad and nearly symmetric, with
a trend for the lines at high n-number to be narrower than the \brg\,
line. With the assumption that the three lines have different optical
depths and consequently trace zones at different physical depths, 
we interprete the observed profiles as evidence that the ionized gas
velocity in the HI emitting region is increasing as we move outwards, as
expected in an accelerating wind more than in an infalling gas.\\
We have modelled the observed line ratios and velocities with a simplified
model for the HI excitation from a circumstellar gas with a velocity law
$V=V_0+(V_{max}-V_0)(1-(r_i/r)^{\alpha})$. Such a comparison indicates 
that the observations are consistent with the emission coming from 
a very compact region of 4-6 R$_{\odot}$, where the gas has been
already accelerated to velocities of the order of 200 \kms, with an associated 
mass flow rate of the {\it ionized} component of the order of
10$^{-7}$ M$_{\odot}$\,yr$^{-1}$. This implies that the observed lines
should originate either from a stellar wind or from the inner part 
of a disk wind, providing that the disk inner truncation radius is close
to the stellar surface. It is also expected that the gas ionization fraction
is relatively high as testified by the high rate of ionized mass loss derived.
Our analysis, however, does not resolve the problem of how to reproduce
the observed symmetrical line profiles, which
at present are apparently difficult to model by both wind and 
accretion models. This probably points to the fact that the real situation 
is more complicated than described in the simple model presented here.}

\maketitle
\section{Introduction}
The emission from hydrogen recombination lines represents the most direct
manifestation of circumstellar activity in young stars. Lines from 
the Balmer series have been the main spectroscopic tool for 
identifying classical T Tauri and Herbig AeBe stars.
 In spite of the large amount of observational data, however, the HI lines are so 
easily excited that it is very difficult to clearly define their origin.
Originally, they were interpreted as being excited in circumstellar 
ionized winds, on the basis of the often observed P Cygni profile 
exhibited by the H$\alpha$ line (Hartmann et al. 1990, Calvet et al. 1992). 
More recently, such an interpretation has been challanged due to the variety 
of the observed Balmer line profiles (e.g. Edwards et al. 1994, Reipurth et al.
1996) that are difficult to reproduce by means of wind models,
and the HI emission has been interpreted also in the framework 
of magnetospheric accretion models (Calvet \& Hartman 1992, 
Muzerolle et al. 1998a). 
Studies of HI lines in pre-main sequence stars have been extended also 
in the near and mid IR (Folha \& Emerson 2001, 
Benedettini et al. 1998, Nisini et al. 1995, Natta et al. 1988).
While the IR HI line ratios are well reproduced by wind models,
they apparently fail to explain the observed profiles of the 
\brg\, and Pa$\beta$ lines. On the other hand, magnetospheric accretion 
models were able to reproduce the \brg\, profiles showing redshifted
absorption components (Muzerolle et al. 1998b) but not all the variety of profiles 
exhibited by T Tauri stars (Folha \& Emerson 2001).
Finally, very recently, spectro-astrometric observations of the Pa$\beta$ line
in four T Tauri stars, have indicated that the emission originates in the
outflowing material (Whelan et al. 2004).\\
While there is ample literature on the origin and observations of HI lines in 
T Tauri and Herbig stars, the interpretation of HI lines in more 
embedded Class I sources, for which optical observations of the Balmer series 
are impeded, is much less defined.
Observations of spectrally-resolved HI lines in low-mass Class I sources 
so far have been limited to the Br$\gamma$ line
(Najita et al. 1996, Davis et al. 2001). At variance with H$\alpha$ 
in T Tauri stars, the profiles
of the \brg\, line appear more symmetric, with no evidence of the typical
P Cygni signature. On this basis Najita et al. (1996) argue against a wind 
origin, while in the framework of magnetospheric accretion models, Muzerolle 
et al. (1998b) use the \brg\, luminosity in a sample of Class I sources to 
infer the mass accretion rates.\\
To better constrain the emission region and the mechanism 
for the excitation of HI lines in Class I sources it is necessary to 
obtain a larger observational set, including both information on 
line profiles and flux ratios among different lines. The aim of this work 
is to start such an analysis using both the ratios from lines in the Brackett 
series (the Brackett decrement) and high resolution spectroscopy of 
different Brackett lines. The source investigated here is \object{HH100 IR}, a 
low mass ($L_{bol} \sim$ 10 L$_{\sun}$, Wilking et al. 1992)  Class I object
located in the R CrA star forming core (D=130 pc, Marraco \& Rydgren 1981). 
This source already
has been recognized as an embedded protostar with high circumstellar 
activity on the basis of its near IR specrum (Greene \& Lada 1996, Nisini et al.
2004), 
and as also suggested by the fact that it is the driving source of an 
Herbig Haro object. It therefore represents a suitable test case to 
perform a detailed analysis of its HI emission.

\section{Observations}
The observations were performed with the ISAAC spectrometer, at the 
VLT UT1 telescope, during 12 and 13 July 2002. 
In the low resolution mode observations we employed a 0\farcs6 slit to acquire 
spectra covering the H and K bands at R$\sim$800. 
In the medium resolution mode, two spectral segments were acquired
centered at 1.629\um\, and at 2.161\um\, and covering 
about 0.6\um\, each. For these observations a 0\farcs3 slit was used, 
giving a resolution of $\sim$8000 and 9000 at 2.1 and 1.6\um, respectively.
For both low and medium resolution observations,
spectra of a standard 
star of B spectral type were obtained at an airmass similar to the
scientific spectra, to correct for telluric absorption and obtain flux
calibrations. The telluric spectra were carefully cleaned of any
intrinsic HI absorption feature before being used.
Wavelength calibrations were performed both using a Xenon lamp 
spectrum taken at the end of the night, and refined each time on
the OH sky lines observed in the spectra. This procedure leads to a
wavelength calibration error of $\sim$ 0.1 \AA\, (i.e. about 1-2 
km\,s$^{-1}$).\\

\begin{table}
\caption[]{Low resolution observations}
\vspace{0.5cm}
\begin{tabular}{cccc}
\hline\\[-5pt] 
 & &  \multicolumn{2}{c}{HH100 IR}\\
  $\lambda$($\mu$m) & Line&  Flux$^{a}$ &$\Delta~F$$^{a}$\\[+5pt]  
\hline\\[-5pt]
1.501 & Br 24 & 3.5 & 0.3\\
1.504 & Br 23 + Mg I & 20.1 & 0.7\\
1.509 & Br 22 & 6.1 & 0.7\\
1.514 & Br 21 & 7.3 & 0.6\\
1.519 & Br 20 & 10.5 & 0.6\\
1.527 & Br 19 & 15.5 & 1.0\\
1.534 & Br 18 & 16.2 & 0.9\\
1.544 & Br 17 & 23 & 1\\
1.556 & Br 16 & 24.8 & 0.9\\
1.571 & Br 15 & 32.5 & 1.0\\
1.588 & Br 14 & 51 & 2\\
1.611 & Br 13 & 64 & 2\\
1.641 & Br 12 & 66 & 2\\
1.681 & Br 11 & 94 & 3\\
1.737 & Br 10 & 155 & 5\\
1.818 & Br 9 & 229 & 26\\
2.166 & Br$\gamma$ & 513 & 17 \\
\hline\\[-5pt]
\end{tabular}

~$^{a}$Fluxes and their errors are expressed in 10$^{-15}$\,erg\,cm$^{-2}$\,s$^{-1}$

\end{table}

\begin{table}
\caption[]{High resolution observations}
\vspace{0.5cm}
\begin{tabular}{cccc}
\hline\\[-5pt] 
 & & \multicolumn{2}{c}{HH100 IR}\\
  $\lambda$ & Line& $V_{LSR}^{a}$ & $\Delta~V$ \\
  ($\mu$m) & & km\,s$^{-1}$ & km\,s$^{-1}$ \\[+5pt]  
\hline\\[-5pt]

1.6114 & Br 13      & -2 & 184 \\
1.6412 & Br 12      & -3 & 190 \\
2.1661 & Br$\gamma$ & -17 & 224 \\
\hline\\[-5pt]
\end{tabular}

~$^{a}$ $V_{LSR}$ corrected for the velocity of the R CrA cloud 
(5.8 \kms, Hariu et al. 1993)

\end{table}

In the low resolution spectra we detected 
the Br$\gamma$ at 2.166\um\, in the $K$ band spectrum, and lines 
from the higher levels of the Brackett series, from Br 9 up to 
Br $24$, in the $H$ band (Figure 1 and Table 1). The Br23 line,
which shows a line flux much higher than the adjacent lines, is
blended with a feature of Mg I.
The medium resolution spectra covered the Br$\gamma$, the Br 12 (1.6113\um) 
and the Br 13 (1.6411\um) lines, which all appear resolved. Figure 2 
shows the profiles of the lines, while in Table 2 we report the 
velocity information (V$_{LSR}$ and $\Delta$V (FWHM) ) derived through a
Gaussian fit. We estimate an error in the line width determination that
is between 5 and 10\kms, based on
the difference between the FWHM of the Gaussian fit with that directly measured 
with respect to the observed peak.
The V$_{LSR}$ velocities are corrected for the systemic velocity of the R CrA 
cloud taken as 5.8 km\,s$^{-1}$ (Harju et al. 1993).

\begin{figure*}[!ht]
\includegraphics[width=16cm]{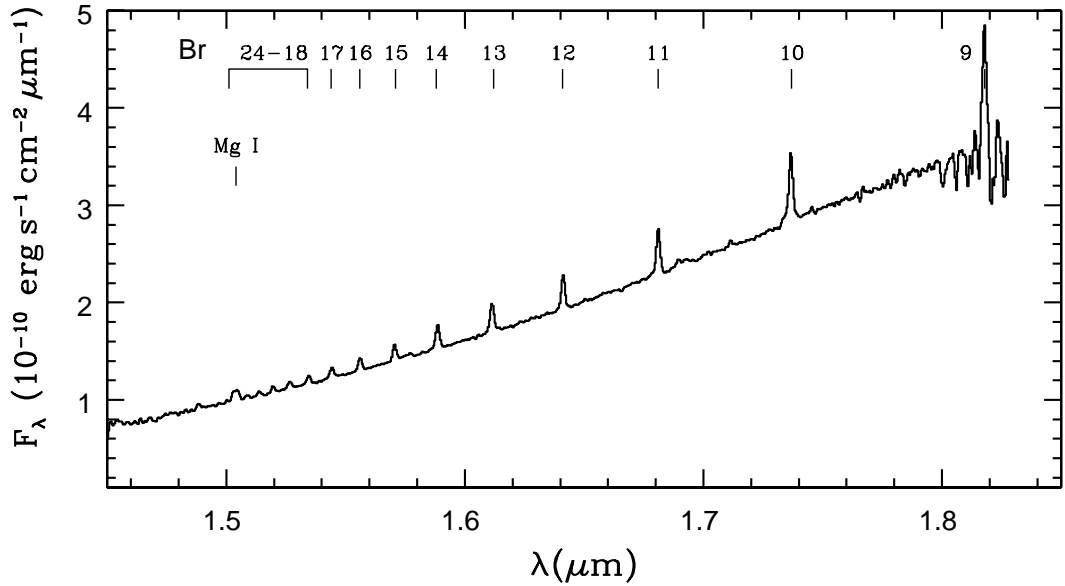}
\caption{Low resolution spectra in the H band of \object{HH100 IR}, 
showing the emission lines from the Brackett series.}
\end{figure*}

\begin{figure}
\resizebox{\hsize}{!}{\includegraphics{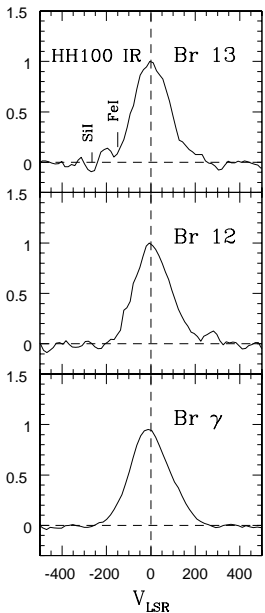}}
\caption{Resolved spectra of the Br$\gamma$, Br 12 and Br 13 lines in \object{HH100 IR}.
The underlying continuum of each spectrum has been subtracted and each line
is normalized to its peak intensity. The V$_{LSR}$
is relative to the stellar velocity (i.e. the velocity of the associated
molecular cloud has been subtracted).}

\end{figure}

\section{Line ratios}
The observations of a large number of HI lines from the Brackett series 
allow a detailed study of the decrement of this series, from which 
information on the line excitation can be derived. In normal HII regions, 
the decrement is actually used to derive the extinction through the 
emitting region, assuming that Case B recombination holds (Hummer \& Storey 1987). 
However, the excitation of NIR HI lines in Young Stellar Object (YSO)
 envelopes usually cannot be 
reconciled with a standard Case B HII region since, due to
the high densities of the source circumstellar gas, the 
lines may remain optically thick up to high-n levels. 
Therefore, to proceed with a correct analysis of the observed line ratios
we first need to correct the observed lines for the reddening.
Estimates on the visual extinction for \object{HH100 IR}, based on the optical depth 
of both the silicate feature at 9.1\um\, and the 3$\mu$m ice band absorption feature,
suggest a value around 25 mag (Whittet et al. 1996). 
Such a value is not affected by the 
IR variability of \object{HH100 IR} (Molinari et al. 1994) since  
Graham (1998) found no variability in the 3$\mu$m ice band over 
a period of 7 yr. In addition, an $A_{\rm v}$ between 20 and 30 mag has been
also estimated from the comparison among the intrinsic and observed 
(H-K) colors, assuming the stellar spectral type derived from the
photospheric features detected in the IR spectrum of the source 
(Nisini et al. 2004). 
Therefore a value of $A_{\rm V}$ =25 mag, together with the
reddening law by Rieke \& Lebofsky (1985), have been adopted 
to correct the observed fluxes for the reddening.
Figure 3 shows the ratios of the different Brackett lines with respect
to the \brg\, line intensity. In the same figure the ratios expected 
for a Case B recombination are also plotted for the extreme cases of 
$T$=6000 K, $n$=10$^{4}$ cm$^{-3}$ and $T$=10000 K, $n$=10$^{6}$ 
cm$^{-3}$, taken from Storey \& Hummer (1995). The observed ratios
are always higher than the Case B values, which indicates that the
Brackett lines remain optically thick up to high values of the n number.
The displacement from the Case B curves could be due to a wrong 
extinction value assumed, but an $A_{V}$ value as low as $\sim$5 mag would 
be needed to reconcile the observations with the Case B recombination.\\
In the figure, we also show the Brackett decrement in the 
case of emission from optically thick ($\tau >>$1) lines originating 
from a region of fixed radius at $T$=10\,000 K. In this extreme case, lines from high
n-number remain always brighter than the \brg.
It is evident from this figure that the observed ratios lie in an 
intermediate situation between these two extreme cases. Since for 
lines of a given series $\tau \propto \lambda$, the optical depth 
also decreases as the n-number increases. As a consequence, assuming that 
each line originates from a surface at which $\tau \sim$ 1, it is expected 
that different lines trace a different emitting region whose size decreases 
as the n-number increases (see e.g. discussion in Benedettini et al. 1999). 

\section{Line profiles}

The profile of the \brg\, line in \object{HH100 IR} is broad (FWHM $\ga$ 200 \kms) and 
nearly symmetric. 
This profile is indeed similar to 
those observed from other Class I sources 
(Najita et al. 1996, Davis et al. 2001) and from many T Tauri objects
(Folha \& Emerson 2001). These previous studies also showed that the
\brg\, line peaks are often blueshifted, as observed also in our case, where
a $V_{LSR}$ of $-$14 \kms\, has been measured. Such blue-shifted symmetrical 
profiles are difficult to reproduce 
 either by wind models and by magnetospheric accretion models. 
Wind models generally predict red-shifted peaks and P-Cygni absorption 
features. Most of the profile calculations are however  done for the
Balmer lines (e.g. Calvet et al. 1992), while predictions for NIR lines 
are seldom found. The Br$\gamma$
and Br$\alpha$ profiles derived from Hartmann et al. (1990) in their
magnetically driven wind model, do indeed lack P-Cygni absorption features
and are more symmetric, with a FWHM of the order of 200 \kms, comparable to
the values observed in Class I sources,
but they still present pronounced asymmetries and they are mostly
red-shifted. In the case of wind models, a blue-shifted centroid may
result from the occultation of the redshifted part of the profile by an
optically thick circumstellar disk, as is observed in optical forbidden
lines (e.g. Edwards et al. 1987). Such an effect however should be 
accompanied by a significant asymmetry in the redshifted side of the 
profile which apparently is lacking here.
Magnetospheric accretion models, on the other hand, tend to predict profiles
with blue-shifted peaks, as in our case, but where such a blue-shift centroid is
the result of strong asymmetries caused by the suppression of the redshifted 
part of the profile (e.g. Muzerolle et al. 1998a).
Recently, Muzerolle et al. (2001) provided  more refined line profile models,
including line damping due to different broading mechanisms, which have the net
effect of filling in the red-shifted absorption component causing a much more
symmetric and centrally peaked profile. This effect is however particularly 
important only for the H$\alpha$ line and becomes less significant for the higher
Balmer and near IR lines. 
More symmetric profiles  of the \brg\, may be predicted at privileged inclination 
angles of the accretion disk (Muzerolle et al. 1998a), but even in the extreme situation of 
$i\sim$70$^\circ$ the predicted red-shifted asymmetries should be 
still recognizable in our high S/N spectrum.
Finally, the absorption of the red-shifted side of the profile also cause the 
FWHM line width to narrow significantly, and indeed the \brg\, profiles shown in 
Muzerolle et al. (1998a, 2001) always have a FWHM of about 100 \kms\, or less,
thus much narrower than the values we observe. Such a discrepancy between
 the FWHM predicted by the magnetospheric accretion models and the observed 
 wider \brg\, line-widths was also pointed out by Folha \& Emerson (2001) for their
 sample of T Tauri observations. Since the line width 
 depends on the infall velocity field in the accretion flow, which in turn
 is a function of the source gravitational potential energy, such a width could be
 larger in sources with a mass greater than the canonical value assumed by the Muzerolle
 et al. models, i.e. 0.5 M$_{\sun}$. In the specific case of \object{HH100 IR}, a stellar
 mass of 0.4 M$_{\sun}$ has been estimated (Nisini et al. 2004), so
 in principle line widths even narrower than those predicted by Muzerolle
 et al. should be expected.
\\

On the basis of these considerations, we believe that at present both the existing wind and
accretion models still fail in reproducing the observed HI near-IR symmetric profiles, 
and therefore it is not possible to favor one of these two excitation
mechanisms on the basis of profiles alone.

The profiles of the observed Br12 and 13 lines also appear broad
but with FWHM velocities smaller than the \brg\,
line. These profiles are also rather symmetric: two apparently 
blue-shifted absorptions in the Br 13 line can be ascribed to 
the presence of two photospheric absorption features, as indicated in 
Figure 3. If we correct for these absorption features, the Br13 line-width
becomes $\sim$190 \kms, i.e. more similar to the Br12 line and 
still significantly lower than the \brg\, line-width.
It seems therefore to be a trend for the 
FWHM velocity to decrease with higher quantum numbers. 
We have seen
that the observed Brackett lines are thick and have different optical depths, 
thus they trace zones at different physical depths in the emitting region.
Since moreover the optical depth decreases with increasing upper quantum number
(i.e. with decreasing wavelength), it is expected that high-n lines 
trace regions more internal than the \brg\, line. Therefore if we assume that
the observed line widths measure the maximum velocity attained at the
emitting region surface, the observed trend
would imply that the velocity is increasing going outwards,
as expected, e.g., in an accelerating wind. With the assumption
that such a velocity trend can be ascribed to the gas kinematical motion alone, 
then the different observed linewidths are incompatible with a velocity law due to any 
accretion process or Keplerian rotation (such as $v \propto r^{-1/2}$) 
where the velocity is decreasing with 
the distance from the central object.
Emission from a flattened geometry 
viewed at different inclination angles cannot produce changes in this trend.
Muzerolle et al. (2001), on the other hand, presented evidence that other line
broadening effects, radiative and Stark broadening in particular, may dominate the
line-width in some circumstances, an effect which is particularly important for
the H$\alpha$ line. However,
The Muzerolle et al. (2001) models show that these broadening effects become 
negligible for near IR lines, and hence the line-widths of the Brackett lines 
are more likely to reflect the dynamical properties of the gas in the emitting region
than other broadening effects.

A rough estimate of the size of the emitting region can be given
assuming that the flux in the different lines is given by a blackbody
emission at T$\sim$10\,000 K, i.e.:
\begin{equation}
F_{\lambda} = \pi\,\frac{R_{\tau=1}^2}{D^2}\,B_{\lambda}(\lambda,T)\lambda\frac{\Delta V}{c}.
\end{equation} 
Taking as $\Delta V$ the widths of the observed lines and a distance of 130 pc,
we derive that the \brg\, emission size is of the order of 4\,10$^{11}$ cm
while the Br12,13 emission sizes are of the order of 2\,10$^{11}$ cm. This
indicates that the HI lines originate from a region of only 4-6 R$_{\sun}$,
thus very close to the protostellar photosphere, which is expected to be
a few solar radii in diameter. 
 Since the emitting region is not vey large, the central source may 
produce an occultation of the redshifted part 
of the flow. A simple way to check the amount of this occultation is through
geometrical considerations. The emission originates from the surface
at which $\tau \sim$1, thus the amount of flux not reaching the observer due
to the central star can be estimated as the ratio between the 
emission region surface and the source projected area. 
This estimate shows that only $\sim$ 5 \% 
of the total line emission would be hidden by the central source, 
and thus difficult to detect
although it should mainly affect the line redshifted wing.
An occultation from a circumstellar disk should in fact produce a 
more severe effect on the line profile. Such an effect is not 
observed also in HI lines from T Tauri stars while it is often present
in other optical forbidden lines such as [OI] (e.g. Edwards et al. 1987,
Takami et al. 2001).
This has been interpreted as evidence for the presence of internal
disk holes allowing the redshifted part of lines coming from compact
regions to reach the observer while obscuring that of the 
forbidden lines, originating from more external and lower density 
regions (Takami et al. 2001).

\section{Comparison with models of ionized envelopes}

There are very few models in the literature treating in detail 
(i.e. with a full radiative transfer treatment and taking into account
geometrical effects) the excitation of HI IR lines and providing
predictions for both flux ratios  and line profiles. Natta et al. (1998)
provide Pa$\beta$, Br$\alpha$ and Br$\gamma$ intensities of T Tauri 
stars in the framework of a partially ionized wind model but without 
any prediction of line profiles. Muzerolle et al. (1998a), on the other hand,
compute the \brg\, profile expected from their magnetospheric accretion 
model, but the line fluxes are derived only for Pa$\beta$ and \brg\, 
lines.\\
Giving the difficulty of comparing
our data with specific models, we have adopted a more general approach, 
considering a simple model of an ionized envelope in which the gas velocity follows
a general law of the type $V=V_0+(V_{max}-V_0)(1-(r_i/r)^{\alpha})$,
which assumes that the gas is accelerated at a maximum velocity 
$V_{max}$ at a distance that depends on the parameter $\alpha$.
The ionized envelope
extends from an initial radius $r_i$ to an outer radius $r_0$. Inside this
region, the electron density follows from the continuity equation and it is
equal to $n_e = \dot{M}\,x_e/4\,\pi\,m_H\,r^2\,V(r)$, where $x_e$ is the 
fractional ionization and $\dot{M}$ is the rate
of mass flow inside the region.  The radiation transfer has
been treated in the Sobolev approximation assuming the gas in LTE and 
following the formalism described in Nisini et al. (1995), where a 
discussion about the limitations of the adopted assumptions is also 
given.\\

\begin{figure*}[!ht]
\sidecaption
\includegraphics[width=12cm]{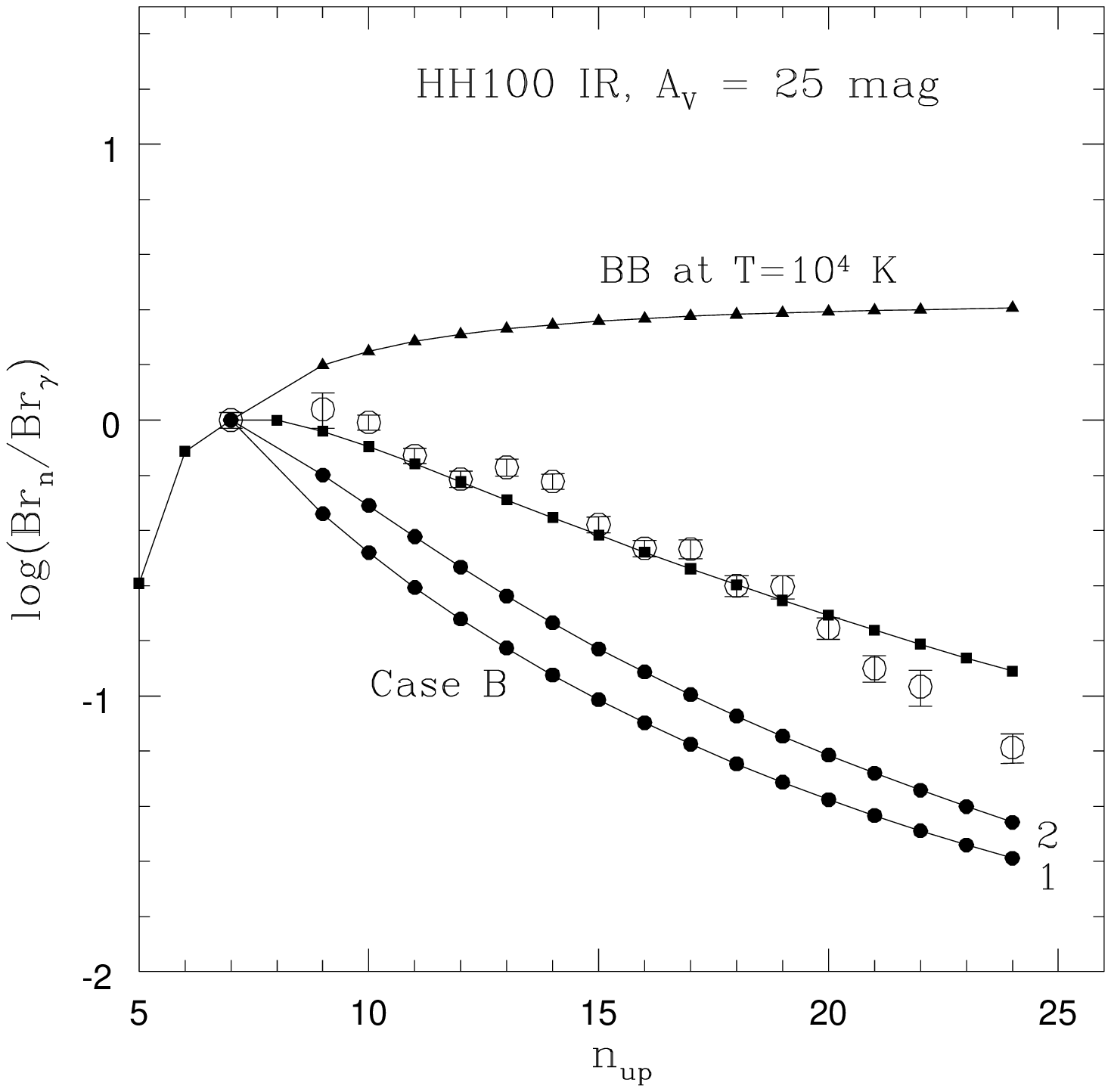}
\caption{Ratio of the Brackett lines with respect to the \brg\, plotted
as a function of the upper quantum number. Open circles are the 
ratios measured on \object{HH100 IR}, dereddened by a visual extinction of 25 mag.
The two curves indicated with filled dots are the ratios expected from
Case B recombination (Hummer \& Storey 1987) assuming {\bf 1-} $T$=6000 K, $n$=10$^{4}$ 
cm$^{-3}$, {\bf 2-} $T$=10\,000 K, $n$=10$^{6}$ cm$^{-3}$. Triangles are the ratios
expected in case of emission from an optically thick region at $T$=10\,000 K.
The squares represent our best fit model to the observed points 
assuming a spherically symmetric model with an accelerating 
velocity law. The parameters for this
model are the following (see text for the parameters definition): 
$r_0$=3 R$_*$ , $\dot{M}_{ion}$=2\,10$^{-7}$ M$_{\sun}$\,yr$^{-1}$, 
V$_0$=30 \kms, V$_{max}$=230 \kms\,
and $\alpha$=4. }
\end{figure*}

We have tuned the input parameters in such a way as to reproduce
the line ratios, line fluxes and the observed line widths with the assumption
that the FWHM traces the maximum velocity attained in the line emitting region.
This last constraint is equivalent to reproducing the 
emitting region sizes as estimated in the
previous section. 

In this framework, the optical depth at line center, which is the main
parameter affecting the line decrement, is given, as a function of the distance
from the initial radius $r_i$,  by:
\begin{equation}
\tau =
\frac{K_{\nu}\,c}{\nu\,r_i^{3}}\,\left(\frac{\dot{M}\,x_e}{4\,\pi\,m_H}\right)^{2}
\,V(r)^{-3}
\end{equation}
where $K_{\nu}$ is the line absorption coefficient and $\nu$ the line center frequency.
From this expression, we see that the optical depth is strongly 
affected, by the adopted velocity law, the initial radius $r_i$ and the rate 
of ionized gas flow
$\dot{M}_{ion}$=$\dot{M}\,x_e$. We have chosen that the emitting region begins at the
 protostellar radius $R_{*}$ , assumed equal to 3\,R$_{\sun}$. This is the predicted 
radius  of protostellar objects and it is consistent with the 
derived value of (3.5$\pm$0.7)\,R$_{\sun}$ estimated from the inferred spectral
type and stellar luminosity (Nisini et al. 2004).
 
 With this condition, to maintain the lines optically thick up to 
high-n values, the $\dot{M}_{ion}$ value needs to be sufficently high, e.g. larger than
10$^{-7}$ M$_{\sun}$\,yr$^{-1}$. At the same time, the absolute line 
fluxes are also sensitive, in addition to  $\dot{M}_{ion}$, to the gas 
temperature and the emitting region size. Assuming $T$= 8000 K, we find that
$r_0$ values of a few stellar radii are needed to mantain the observed line 
fluxes close to the observed absolute values.
Colder emission may be consistent with slightly larger emission regions, which
however cannot exceed $\sim$8 $R_*$ for $T \ga$ 4000 K.

In Fig. 3 we show the predicted decrement in our best fit model with the following
parameters:  $\dot{M}\,x_e$=2\,10$^{-7}$ M$_{\sun}$\,yr$^{-1}$, $r_0$=3\,R$_{*}$, 
V$_0$=30 \kms, V$_{max}$=230 \kms\,
and $\alpha$=4. Such a model also reproduces fairly  well the dereddened
flux of the \brg\, line (e.g. 5.9 10$^{-12}$ erg\,cm$^{-2}$\,s$^{-1}$ vs
(6.2$\pm$.2)\,10$^{-12}$ erg\,cm$^{-2}$\,s$^{-1}$); moreover it predicts that the 
maximum velocities corresponding to the radii where the \brg,
Br12 and Br13 lines attain $\tau$=1 are  220, 190 and 180 \kms\, respectively,
also in agreement with the observed line widths. The model however 
overestimates the emission from high n$_{up}$ numbers with respect to the
observed values. The high n$_{up}$ lines originate
from inner regions where, according to the adopted velocity law, the speed 
is still low; consequently, 
the expected optical depth of these lines is mantained high despite their
 higher frequency, according to the relationship (2).  
A better fit to the data can be obtained by assuming a different velocity 
law with a much higher initial velocity ($\sim$ 180 \kms), which however 
seems unphysical implying that the ionized gas is flowing from the stellar 
surface already at a velocity close to the escape speed.

In summary, and considering the oversemplification of the adopted model,
we can reasonably conclude that the different observing features can be
accounted for by an ionized gas, emitted in a region of a few 
stellar radii close to the source where the gas is already accelerated at
about 200 \kms\, and it is expanding, after an initial steep
acceleration, with a rate
of (ionized) mass loss of the order  of  10$^{-7}$ M$_{\sun}$\,yr$^{-1}$.\\
In deriving this mass loss rate value, it is assumed that the ionization
fraction does not change over the line emitting region. While this may 
not be adequate if one considers the wind ionization structure 
due to the stellar irradiation alone (e.g. Natta et al. 1988), 
it is a rather acceptable approximation in the present case,
given the fact that the lines originate in a compact emitting region close
to the source, and that 
other heating mechanisms have been recognized to be important
in maintaining a high degree of ionization in collimated winds 
(e.g. Shang et al. 2002).

The origin of such an emission in a very compact region close to the 
source suggests excitation from a stellar wind, more than from a
disk-wind unless the emission comes from the inner section of a 
disk extending down to the source surface. This evidence naturally
excludes an origin in MHD disk-winds (e.g. Ferreira
\& Pelletier 1993)  where the disk is 
truncated at large distances from the source by the action of a strong 
magnetic field. 

The  $\dot{M}_{ion}$ value derived in our model is moreover large 
enough to suggest that the gas ionization in the emitting region should be 
high. An ionization fraction of $\sim$ 0.1 would 
imply an $\dot{M} \sim$ 10$^{-6}$M$_{\sun}$\,yr$^{-1}$. 
Values of the order of 10$^{-7}$--10$^{-8}$\, M$_{\sun}$\,yr$^{-1}$ 
have been estimated in T Tauri stars (see e.g. Natta et al. 1988), it is 
however expected that the efficiency of the mass loss mechanism is
higher for younger sources.\\
Finally the adopted velocity law impies that in the outflowing gas 
 the velocity rapidly increases with the distance from the central source and
it is already high ($\sim$200\kms) at about two stellar radii. This velocity
trend, which is consistent with magnetically-driven stellar winds
(e.g.  Lago 1984, Hartmann 1990) as well as with magnetocentrifugal
wind models like X-winds (Shu et al. 1994) has been also measured in some
T Tauri stars from spectro-astrometric observations of the H$_{\alpha}$ 
emission (Takami 2001, 2003). 

 Our simplified model, in which LTE emission is assumed, necessarily
predicts symmetric and centrally peaked emergent line profiles. The
 P Cygni profiles and blue-shifted absorption components which seem to be 
 a common characteristics of wind models thus far explored, result when large 
 line opacities are coupled with significant deviations from equilibrium
in the external expanding region of the wind, where the increasing inefficiency
of collisions causes a  decrease of the source function. 
On the other hand, the mere fact that the observed lines appear symmetric and
almost Gaussian could be an indication that LTE conditions for the Brackett lines
actually hold in the region where they originate. 
Indeed, Nisini et al. (1995) compared their LTE model with models treating in detail
the hydrogen level populations, reaching the conclusions that the LTE assumption
is a reasonably good approximation of the Brackett lines, while it may be 
not totally adequate for the Pa$\beta$ line.
Moreover, the limited \brg\, line profile predictions presented by Hartmann (1990)
shows that since the upper hydrogen levels reach equilibrium 
more easily than the Balmer lines, the resulting absorption features 
are much weaker and the degree of asymmetry significantly reduced. 
Obviously a more refined model than the one presented here, with a full 
statistical equilibrium treatment and a more realistic geometry,
is needed to better explore the ability of wind models to reproduce the 
observed Brackett symmetric profiles.

\section{Conclusion}

The analysis of the Brackett decrement in the \object{HH100 IR} Class I source 
has shown that the observed Brackett lines remain optically thick
up to a high quantum number. In addition, the profiles of the \brg, Br 
12 and Br 13 lines are fairly symmetric with a tendency of the line 
FWHM to decrease as the n-number increases. On the basis of this 
evidence we argue that a wind-like mechanism, where the gas is 
accelerating outwards is  more suited to reproduce the observed 
features. A comparison with a very simple model suggests 
that the emission region should be very compact and close to the 
stellar surface, a fact favouring a stellar wind or the inner region of
a disk wind with a small truncation radius as the natural emission 
region. The mass loss rate of the ionized gas should be fairly high, of the order of 
$\dot{M}_{ion} \sim$ 10$^{-7}$M$_{\sun}$\,yr$^{-1}$, to maintain the 
lines optically thick. This suggests that the ionization fraction 
in the emitting gas should not be smaller that $\sim$0.1 to have a 
total mass loss rate not exceeding typical values expected from Class 
I objects.\\
Our analysis, however, does not solve the problem of how to reproduce 
the observed symmetrical line profiles of optically thick IR lines, 
which at present are apparently difficult to modell by both wind and 
accretion models. This probably points to the fact that the real situation 
is more complicated than described in the simple model presented here.
Another caveat to the derived conclusions is the near 
IR photometric variability of \object{HH100 IR}, which shows differences in 
the $K$ band of more than a magnitude over a period of a few years. 
Such a variability can also affect the line emission of this source.
Indeed, variability not only of HI NIR line fluxes but also of their 
 profiles has been observed in other Class I sources exhibiting 
photometric variability (Nisini et al. 1994). \\
Finally,  our conclusions are valid for the specific case of HH100IR and
as such cannot  be generalized to other Class I objects. Indeed, more
asymmetric profiles of the \brg\, line in similar sources have been 
succesfully modelled by magnetospheric accretion models, such as e.g.
WL16 (Muzerolle et al. 1998b). For a  better understanding of the origin of 
the HI lines in
Class I sources, a more detailed modelling effort is needed, allowing
us to fit in a consistent way both the intensity and profiles of the 
different observed lines.

\begin{acknowledgements}
We thank the anonymous referee for making useful suggestions which helped
to clarify some aspects of the paper.
This research has made use of NASA's Astrophysics Data System Bibliographic
Services and the SIMBAD database, operated at CDS, Strasbourg, France.
\end{acknowledgements}


\begin{thebibliography}{}

\bibitem{} Benedettini, M., Nisini, B., Giannini, T. et al. 1998, A\&A, 339, 159
\bibitem{} Calvet, N. \& Hartmann, L. 1992, ApJ, 386, 239
\bibitem{} Calvet, N., Hartmann, L. \& Hewett, R. 1992, ApJ, 386, 229
\bibitem{} Davis, C.J., Ray, T.P., Desroches, L. \& Aspin, C. 2001, 
MNRAS, 326, 524
\bibitem{} Edwards, S., Cabrit, S., Strom, S.E., Heyer, I., Strom, K.M. \&
Anderson, E. 1987, ApJ, 321, 473
\bibitem{} Edwards, S., Hartigan, L., Ghandour, L. \& Andrulis, C. 1994, AJ, 108, 1056
\bibitem{} Ferreira, J. \& Pelletier, G. 1993, A\&A, 276, 625
\bibitem{} Folha, D.F. \& Emerson, J.P. 2001, A\&A, 365, 90
\bibitem{} Graham, J.A. 1998, ApJ, 492, 213
\bibitem{} Greene, T.P. \& Lada, C.J. 1996, ApJ, 112, 2184
\bibitem{} Harju, J., Haikala, L.K., Mattila, K., Mauersberger, R., Booth, R.S.,
Nordh, H.L. 1993, A\&A, 278, 569
\bibitem{} Hartmann, L., Calvet, N., Avrett, E.H. \& Loeser, R. 1990, 
ApJ, 349, 168
\bibitem{} Lago, M.T.V.T. 1982, MNRAS, 198, 445
\bibitem{} Hummer, D.G. \& Storey, P.J. 1987, MNRAS, 224, 801
\bibitem{} Marraco, H.G., \& Rydgren, A.E. 1981, AJ, 86, 62
\bibitem{} Molinari, S., Liseau, R., Lorenzetti, D. 1993, A\&AS, 101,59
\bibitem{} Muzerolle, J., Calvet, N., \&  Hartmann, L. 2001, ApJ, 550, 944
\bibitem{} Muzerolle, J., Calvet, N., \&  Hartmann, L. 1998a, ApJ, 492, 743  
\bibitem{} Muzerolle, J., Hartmann, L. \& Calvet, N. 1998b, AJ, 116, 
2965
\bibitem{} Najita, J., Carr, J.S. \& Tokunaga, A.T. 1996, ApJ, 456, 
292
\bibitem{} Natta, A., Giovanardi, C., Palla, F. 1988, ApJ, 332, 921
\bibitem{} Nisini, B., Antoniucci, S., Giannini, T. \& Lorenzetti D. 2004, in
preparation
\bibitem{} Nisini, B., Milillo, M., Saraceno, P., Vitali, F. 1995, A\&A, 302, 191
\bibitem{} Nisini, B., Smith, H.A., Fischer, J., Geballe, T.R.
1994, A\&A, 290, 463
\bibitem{} Reipurth, B., Pedrosa, A. \& Lago, M.T.V.T 1996, A\&ASS, 120, 229
\bibitem{} Rieke, G.H. \& Lebofsky, M.J. 1985, ApJ, 288, 618
\bibitem{} Shang, H., Glassgold, A.E., Shu, F.H. \& Lizano, S. 2002, ApJ, 564, 853
\bibitem{} Shu, F.H., Najita, J., Ruden, S.P., \& Lizano, S. 1994, ApJ, 429, 797
\bibitem{} Storey, P.J. \& Hummer, D.G. 1995, 272, 41
\bibitem{} Takami, M., Bailey, J., Gledhill, T.M., Chrysostomou, A., 
Hough, J.H. 2001, MNRAS, 323, 177
\bibitem{} Takami, M., Bailey, J., \& Chrysostomou, A. 2003, A\&A, 397, 675
\bibitem{} Wilking, B.A., Greene, T.P., Lada, C.J., Meyer, M.R., Young, E.T.
1992, ApJ, 397, 520
\bibitem{} Whelan, E.T., Ray, T.P., \&Davis, C.J. 2004, A\&A, 417, 247
\bibitem{} Whittet, D.C.B., Smith, R.G., Adamson, A.J. et al. 1996, ApJ,
458, 363
\end{thebibliography}
\end{document}